# Experimental measurements and modeling of characteristic time scales in single iron particle ignition

Liulin Cen[a,b], Yong Qian[a,*], XiaoCheng Mi[b,c], Xingcai Lu[a]

[a]Key Lab. for Power Machinery and Engineering of M. O. E, Shanghai Jiao Tong University, 200240, Shanghai, P. R. China
[b]Department of Mechanical Engineering, Eindhoven University of Technology, Eindhoven, The Netherlands
[c]Eindhoven Institute for Renewable Energy Systems, Eindhoven University of Technology, Eindhoven, The Netherlands

___________________________________________________________________

**Abstract**

Recyclable metal fuels such as iron are promising carbon-free energy carriers for heat and power. In such systems, particle ignition characteristics strongly affect combustion efficiency and combustor stability, making them critical for burner and reactor design. However, predictive ignition modelling remains limited by the lack of time-resolved data for single-particle solid-phase oxidation and phase transitions. In this work, digital in-line holography combined with ultra-high-speed single-color pyrometry is used to resolve characteristic solid-phase oxidation times of spherical micron-sized iron particles burning in well-defined hot oxidizing environments. Three temperature plateaus are identified, corresponding to FeO melting, the $\gamma$-Fe to $\delta$-Fe transition, and Fe melting, from which pre-melting oxidation times and melting durations are extracted. An ignition model based on solid-phase iron oxidation kinetics following a parabolic rate law, coupled with external-oxygen-transport-limited description, is used to simulate these characteristic times. The model accurately captures the FeO-scale pre-melting oxidation time, which is nearly independent of oxygen concentration, while the FeO, $\gamma$-Fe to $\delta$-Fe, and Fe melting stages show strong oxygen-concentration dependence consistent with external-oxygen-transport-limited reaction rates. These measurements and simulations provide the first diameter-resolved dataset for FeO and Fe melting processes and show that this modelling framework can quantitatively predict characteristic times for single iron particles in metal-fuel applications.

___________________________________________________________________

*Corresponding author.

**Novelty and significance statement**

This work presents a time-resolved description of single iron-particle ignition structured around FeO-involving phase-change processes, rather than a single global solid-oxidation-time. It provides the first diameter-resolved experimental benchmarks of FeO-scale melting and iron melting time scales for individual iron particles and uses these phase-resolved metrics to rigorously assess a solid-phase oxidation modeling framework. The ability of the model to quantitatively reproduce all experimentally measured characteristic times without empirical tuning establishes significantly strengthens confidence in physics-based ignition models for the design and optimization of iron-fuel reactors and metal-fuel energy systems.

1. **Introduction**

Recyclable metal fuels have attracted growing interest because of their high energy density, convenient storage and transport, and essentially carbon-free combustion. Among them, micron-sized iron particles are particularly promising owing to their abundance, environmental compatibility, and the ease of recycling their oxides [1]. After combustion, the oxides can be reduced back to iron using renewable electricity or hydrogen, enabling a closed metal fuel cycle for safe, zero-carbon storage and transport of renewable energy [2].

In such a cycle, the ignition behavior of iron particles directly affects combustion efficiency and stability and is therefore a key combustion characteristic. For a single particle, it is commonly described by the ignition temperature and ignition delay time. Several independent studies have measured the ignition temperature of single spherical iron particles in hot environments [3–6], defining it as the critical ambient temperature at which thermal runaway occurs, with reported values of 1100–1200 K. These values agree well with the predictions from the model of Mi et al. [7], indicating good predictive performance for ignition temperature. However, agreement in ignition temperature alone does not guarantee an accurate description of the full ignition time scales.

In addition to the ignition temperature, the ignition delay time is also a key quantity for iron particle ignition, but measurements for single particles are scarce and its definition is not uniform. Panahi et al.

[8] defined ignition delay as the time from particle entry into the hot environment to the first luminosity recorded by a high-speed camera, whereas Li et al. [9] defined it as the time from the first luminous appearance to complete melting. Both definitions depend on when luminosity is first detected and are therefore sensitive to the optical setup, which complicates comparison between experiments. A more physical definition based on the time from particle entry to thermal runaway would avoid this issue, but at the corresponding temperatures (~1100-1200 K) the visible and near-infrared emission is too weak for current radiation-based diagnostics to be used reliably.

Ning et al. [10] instead introduced the solid-oxidation-time (SOT), defined as the time from when an iron particle enters a high-temperature oxidizing environment until the iron melts. Since the melting point of iron is fixed, SOT is insensitive to differences in optical diagnostic setups. However, during SOT the particle undergoes several multiple successive stages (heating to FeO melting, FeO melting, further oxidation with molten FeO, and Fe melting), which introduces additional complexity and uncertainty to model validation. Nguyen et al. [11] simulated iron particle combustion using a solid-phase iron oxidation kinetics model following a parabolic rate law [7] and a first order surface kinetics model [12,13]. Both models underestimate SOT and fail to reproduce its dependence on oxygen concentration compared to experimental results [10]. Awad [14] further extended the parabolic rate law model by incorporating Knudsen-transition transport [15,16] and surface chemisorption of $O_2$ into SOT, but the simulations still failed to match the experiments quantitatively. Although SOT is a compact descriptor, its multi-stage nature still complicates rigorous theoretical validation.

Further experimental refinement and conceptual simplification of iron particle ignition on the time scale are still needed to clarify the underlying oxidation mechanisms. Many studies [9,10,17–22] infer iron melting from plateaus in particle temperature or luminosity, whereas simulations [11,23–27] predict a brief FeO melting stage lasting only a few hundred microseconds. Before FeO melts, the phase of the particle does not change, so focusing on the oxidation time in this stage, $SOT_{FeO}$, can greatly simplify analysis of ignition. The oxidation rate before FeO melting is expected to follow the parabolic rate law model [7], so experimental determination of $SOT_{FeO}$ and FeO melting times would provide a stringent test and possible refinement of such models. At present, however, only a few studies have shown qualitative experimental schematic of FeO melting [5,28], and direct, time-resolved measurements of $SOT_{FeO}$ and FeO melting times are essentially lacking, which limits quantitative validation and further development of related models. Building on the above considerations, this work addresses the lack of key time-scale data for constraining solid-phase oxidation models by: (i) for the first time resolving and measuring several characteristic oxidation times of single micron-sized iron particles before and after FeO scale melting under well-defined temperature and oxygen concentration conditions; (ii) employing a combined diagnostic scheme of digital in-line holography and ultra–high-speed single-color pyrometry to obtain simultaneous, high spatial and temporal resolution measurements of particle diameter and temperature; and (iii) systematically comparing these characteristic times with simulations based on the solid-phase iron oxidation kinetics model following a parabolic rate law coupled with an external-oxygen-transport-limited description, thereby providing new validations for future numerical simulations of iron particle solid-phase oxidation.

## 2. Experimental setup and methodology

### 2.1 The burner

In this work, a Hencken-type burner fired with methane is used to provide a stable hot environment for iron particle combustion. The particles are ignited by the hot gas, and diameter-resolved characteristic oxidation times near iron melting are measured for single particles using multiple optical diagnostics. The principles and operating parameters of the Hencken-type burner and the particle feeding system have been described in detail in ref. [29].

The schematic of the experimental setup is shown in Fig. 1(a). A stainless-steel capillary at the center of the burner delivers the iron particles. The central tube has an outer diameter of 0.7 mm and an inner diameter of about 0.5 mm, and extends approximately 2 mm above the burner surface. After dispersion by a particle feeding device, the iron particles are carried by an $O_2/N_2$ flow through the central tube into the hot environment. The central tube replaces the premixed methane flame, so the gas temperature at the central tube exit is slightly lower than that of the surrounding region. As shown in previous studies [3,30], the gas above the capillary reaches the same temperature as the surrounding flow after a few millimeters of mixing.

To obtain a more uniform temperature distribution above the central tube, a 455 nm continuous-wave (CW) laser diode (JD60W, Gaomi City Jindiao Industrial Automation Equipment Co., Ltd.), originally designed for laser engraving, is used to directly heat the central tube. The laser spot is about 0.45 mm × 0.50 mm and is fully blocked by the central tube, so that the laser does not irradiate the particles and then affect their ignition. The laser power, measured by a laser power meter (Thorlabs, S146C), was adjusted between 0.47 and 2.70 W depending on the operating condition. In general, higher burner surface or gas temperatures required lower laser power due to stronger heating and reduced heat loss of the center tube, whereas lower-temperature conditions required higher laser power.

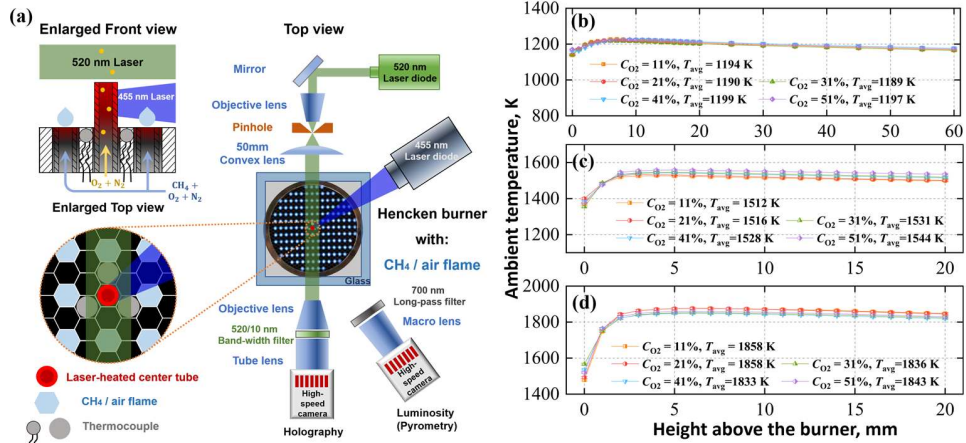

Fig 1. (a): Schematic of the experimental setup; the yellow dot in the enlarged front view marks a single iron particle. (b–d): Centerline gas temperature above the central tube for three ambient conditions, hereafter denoted by their mean values of 1200, 1500, and 1800 K. The temperature at 0 mm is that at the tube exit.

Table 1
Gas flow settings for the different experimental conditions.

| Temperature, K | Gas species | | Oxygen concentration | | | | |
|---|---|---|---|---|---|---|---|
| | | | 11% | 21% | 31% | 41% | 51% |
| 1200 K | Outside the tubes, SLPM | $O_2$ | 1.50 | 3.22 | 4.90 | 5.50 | 7.50 |
| | | $N_2$ | 12.97 | 11.22 | 9.40 | 7.60 | 6.00 |
| | Inside the tubes, SLPM | $O_2$ | 1.60 | 1.60 | 1.60 | 2.65 | 2.50 |
| | | $N_2$ | 0.60 | 0.60 | 0.60 | 0.60 | 0.60 |
| | | $CH_4$ | 0.600 | 0.600 | 0.600 | 0.600 | 0.610 |
| 1500 K | Outside the tubes, SLPM | $O_2$ | 1.30 | 2.65 | 3.00 | 4.00 | 5.25 |
| | | $N_2$ | 9.33 | 7.87 | 6.45 | 5.16 | 3.63 |
| | Inside the tubes, SLPM | $O_2$ | 1.60 | 1.60 | 2.60 | 3.00 | 3.00 |
| | | $N_2$ | 0.80 | 0.80 | 0.80 | 0.80 | 0.80 |
| | | $CH_4$ | 0.695 | 0.695 | 0.700 | 0.700 | 0.710 |
| 1800 K | Outside the tubes, SLPM | $O_2$ | 0.80 | 1.50 | 2.00 | 2.50 | 3.40 |
| | | $N_2$ | 6.15 | 5.01 | 4.19 | 3.59 | 2.57 |
| | Inside the tubes, SLPM | $O_2$ | 2.10 | 2.50 | 3.20 | 3.80 | 4.00 |
| | | $N_2$ | 1.20 | 1.20 | 1.20 | 0.60 | 0.40 |
| | | $CH_4$ | 0.840 | 0.830 | 0.830 | 0.830 | 0.840 |
| Carrier gas, SCCM | | $N_2$ | 2.67 | 2.37 | 2.07 | 1.77 | 1.47 |
| Carrier gas, SCCM | | $O_2$ | 0.33 | 0.63 | 0.93 | 1.23 | 1.53 |

To investigate iron particle combustion under different conditions, three ambient gas temperatures (~1200, 1500, and 1800 K) are used, combined with five oxygen mole fractions from 11% to 51%. Centerline gas temperatures above the central tube, measured with a radiation-corrected type-B thermocouple, are shown in Figs. 1(b–d), and reach a quasi-steady value about 2 mm above the tube exit. The gas flow settings are listed in Table 1.

Three K-type thermocouples are mounted near the burner surface around the central tube to measure the burner surface temperature and thus better characterize the gas temperature experienced by the particles as is shown in Fig 1(a). In the present Hencken-type burner configuration, the ratio of fired to unfired honeycomb cells is 1:3, and the central tube replaces one fired cell. The three K-type thermocouples replace three unfired cells, so that the local ratio of fired to unfired cells around the capillary, and hence the local oxygen concentration, is preserved. The resulting mean burner surface temperatures are listed in Table 2.

Table 2
Burner surface temperatures, K.

| $C_{O2}$ / $T_{gas}$ | 11% | 21% | 31% | 41% | 51% |
|---|---|---|---|---|---|
| 1200 K | 594.4 ± 5.5 | 601.6 ± 9.4 | 610.2 ± 7.8 | 485.9 ± 7.2 | 509.2 ± 6.4 |
| 1500 K | 718.0 ± 10.1 | 730.0 ± 12.2 | 626.3 ± 13.3 | 575.7 ± 15.2 | 587.0 ± 10.8 |
| 1800 K | 843.6 ± 24.5 | 814.3 ± 23.1 | 731.6 ± 24.8 | 749.4 ± 25.8 | 771.0 ± 22.5 |

*2.2 Digital in-line holography*

Digital in-line holography (DIH) is used to measure the diameter of each iron particle entering the burner. Since holographic images can be numerically reconstructed at different object distances ($Z_{obj}$), this technique provides a better depth of field (DOF) than conventional optical imaging.

The holography setup is shown in Fig. 1(a). A CW laser diode (ML3072-520-1W-12T, Shenzhen Julan Optoelectronics Technology Co., Ltd.) emits a collimated beam with a central wavelength of 520 nm. The beam passes through a spatial filter (M-900, Newport) consisting of a 10× microscope objective and a 25 μm pinhole, and is then collimated by a plano-convex lens with a focal length of 50 mm (GLH12-050-050-VIS, Hengyang Optics). The output power of the laser diode is about 318 mW.

The collimated beam illuminates iron particles from the central tube, and the resulting interference pattern is imaged by a 20× microscope (Laowa Aurogon FF 10-50× NA 0.5 Supermicro APO, Anhui Changgeng Optics Technology Co., Ltd.) onto a high-speed camera (FASTCAM mini AX200, Photron). A band-pass filter with a central wavelength of 520 nm and a full width at half maximum of 10 nm (HNIF-010-520-D12, Hengyang Optics) is placed between the objective and the tube lens to suppress out-of-band and stray light. The camera operates at 20,000 fps with an exposure time of 1.05 μs to avoid motion blur. The image resolution is set to 768×336 pixels, corresponding to a field of view of approximately 0.77×0.34 mm$^2$, with the lower edge of the field of view aligned with the central tube exit. The pixel size in object space is 1 μm.

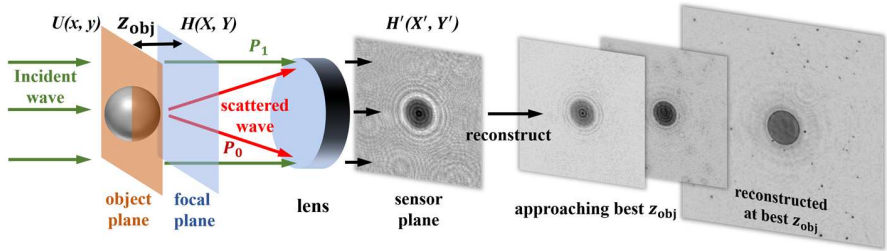

Fig. 2. Schematic illustration of the digital in-line holography principle, where $Z_{obj}$ denotes the distance between the object plane and the focal plane, adapted from ref. [32].

The principle of DIH has been described in detail in refs. [31,32] and is only briefly summarized here. As illustrated in Fig. 2, the collimated laser beam illuminates a spherical particle. The scattered light ($P_0$) from the particle interferes with the unscattered reference field ($P_1$), forming a hologram ($U(x,y)$) that contains both amplitude and phase information. Since the high-speed camera sensor records only intensity, the measured signal at the focal plane is the hologram intensity:

$$H(X,Y) = |U_{\text{detector}}(X,Y)|^2 \quad (1)$$

The hologram $H(X,Y)$ is magnified by the lens to $H'(X',Y')$, which is the image actually captured by the camera. Since $H'$ differs from $H$ only by a constant magnification factor, the following discussion is expressed in terms of $H$ for simplicity.

The hologram $H(X,Y)$ is first background-corrected and normalized to obtain $H_0(X,Y)$, in order to remove the influence of the background on the reconstructed fringes. The reference field $P_1$ is then multiplied by $H_0(X,Y)$, and the hologram is back-propagated using the Fresnel–Kirchhoff diffraction integral [32] to recover $U(x,y)$:

$$U(x,y) = \frac{i}{\lambda} \iint H_0(X,Y) \cdot \frac{\exp(-ik|\vec{r}-\vec{R}|)}{|\vec{r}-\vec{R}|} dXdY \quad (2)$$

where

$$|\vec{r}-\vec{R}| = \sqrt{(x-X)^2 + (y-Y)^2 + Z_{obj}^2} \quad (3)$$

$Z_{obj}$ represents the distance between the object plane and the focal plane. In Eq. (2), $P_1 = 1$ is assumed to simplify the calculation. Eq. (2) is solved using the angular spectrum method; further details are given by Latychevskaia et al. [32]. In this formulation, no additional approximations are introduced. Ref. [32] showed that the radial resolution of the reconstructed hologram is identical to the in-focus optical resolution of the imaging system, while the axial resolution is $\lambda/(NA)^2$, where $NA$ is the numerical aperture. These resolution estimates assume that the sensor records all interference fringes and that noise is negligible. In practice, the finite sensor size and signal noise usually violate this condition, so the effective resolution of DIH is lower than the theoretical limit.

Using the same optical layout and settings mentioned above, holograms of a USAF 1951 resolution target were recorded and reconstructed at different $Z_{obj}$ to evaluate the radial resolution. The results are shown in Fig. 3. At best focus, the system can resolve line pairs with a line width of 1.1 μm. When the object is moved away from the focal plane up to 800 μm, line pairs with a width of 1.23 μm are still resolved. The minimum contrast of the USAF target, $C_{min}$, defined as the minimum contrast between adjacent bright ($I_{light}$) and dark ($I_{dark}$) bars, is given by:

$$C_{\min} = \frac{I_{\text{light}} - I_{\text{dark}}}{I_{\text{light}} + I_{\text{dark}}} \quad (4)$$

As shown in Fig. 3, for the smallest resolvable line pairs, $C_{\min}$ is always greater than 0.3. These results indicate that, in the present setup, the holography imaging system provides a spatial resolution of about 1.23 μm for object distances up to 800 μm away from the focal plane.

In the experiments, the focal plane of the lens was intentionally placed 300–400 μm away from the center of the central tube to enlarge the fringe spacing and facilitate fringe recognition and hologram reconstruction. The inner diameter of the central tube is about 500 μm, so particles at the exit can be located within ±250 μm of the central tube's center plane. Thus, the maximum $Z_{\text{obj}}$ is approximately 650 μm, over which the holographic method still maintains a resolution of about 1.23 μm.

Strong temperature gradients near the central tube may deflect the laser beam and distort the holograms. To assess this effect, a type-B thermocouple with a diameter of about 0.1 mm was placed right above the central tube, and its apparent diameter was measured with holography both without flame and with a ~1500 K flame, in order to evaluate the influence of the local temperature gradient on the reconstruction.

| distance from focal plane | 0 μm | 100 μm | 200 μm | 400 μm | 800 μm |
|---|---|---|---|---|---|
| reconstructed USAF1951 resolution board | | | | | |
| minimum distinguishable element and it's line width | Group 8, Element 6, 1.10 μm | Group 8, Element 5, 1.23 μm | Group 8, Element 5, 1.23 μm | Group 8, Element 5, 1.23 μm | Group 8, Element 5, 1.23 μm |
| minimum vertical contrast horizontal | 0.49 / 0.33 | 0.32 / 0.57 | 0.42 / 0.62 | 0.30 / 0.37 | 0.56 / 0.36 |

Fig. 3. Reconstructed holograms of the USAF 1951 resolution target at different distances from the focal plane and the corresponding spatial resolution. The minimum contrast is defined for the smallest resolvable adjacent line pair; images are shown with inverted grayscale for clarity.

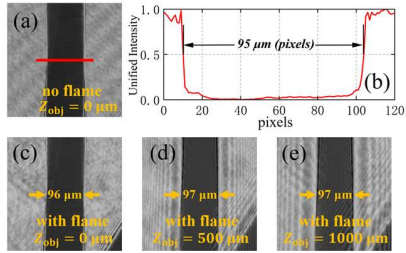

Fig. 4. Reconstructed holograms of the thermocouple wire. (a) Room-temperature image at zero object distance. (b) Normalized intensity along the red line in (a), giving a diameter of 95 μm. (c–e) Reconstructed images at different $Z_{\text{obj}}$ with flame heating (gas temperature ~1500 K).

As shown in Fig. 4(a, b), the thermocouple wire diameter is 95 μm without flame heating. Figs. 4(c–e) show that, with flame heating, the measured diameter increases by 1, 2, and 2 μm at $Z_{\text{obj}}$ of 0, 500, and 1000 μm, respectively. The type-B thermocouple consists of two Pt–Rh wires containing 30% Rh and 6% Rh. Pt–Rh alloys with 0%, 12%, and 20% Rh exhibit linear expansions of 1.34%, 1.36%, and 1.40% at 1500 K relative to room temperature [33]. It is therefore reasonable to assume an expansion of 1.4% for the wire imaged in Fig. 4(a), giving a wire diameter of 96 μm at 1500 K. This agrees with the reconstructed holographic diameters of 96–97 μm at $Z_{\text{obj}}$ of 500 and 1000 μm, demonstrating that beam deflection due to temperature gradients near the central tube exit does not introduce significant errors in the holographic measurements.

It should be noted that the above calibration holograms were reconstructed by manually selecting the reconstruction distance to obtain the sharpest images. This is only possible when the object distance from the focal plane is known. In the experiments, iron particles exit randomly from the central tube, so their $Z_{\text{obj}}$ are not predetermined. The best distance is found by searching for the reconstruction plane that maximizes the intensity gradient at the particle edge. The particle region and the background in the reconstructed image are separated using a k-means clustering algorithm [34].

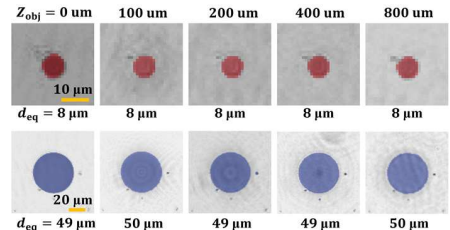

Fig. 5. Reconstructed holograms of two circular targets at different object to focal plane distances ($Z_{\text{obj}}$). The target regions are highlighted in red and blue.

In addition, the DIH setup was used to record holograms of circular targets with known diameters at different object distances in order to assess the accuracy of size measurements for spherical particles, as shown in Fig. 5. At best focus, the measured diameters of the two targets are 8 μm and 49 μm. The results show that, within a defocus range of 800 μm, the system maintains high accuracy for circular diameters, with deviations of at most one pixel. All hologram reconstruction and particle detection procedures were implemented using in-house MATLAB codes.

*2.3 Single-color pyrometry*

Phase-change times such as FeO and Fe melting are captured by directly recording blackbody radiation from the iron particles. A high-speed camera (FASTCAM mini AX200, Photron) equipped with a macro lens (Micro-NIKKOR 55 mm f/2.8 Ai-S, Nikon) is used. A long-pass filter (LP700, PHTODE) with a cutoff wavelength of 700 nm is mounted in front of the lens to block laser light and methane flame emission. The camera exposure time is fixed at 1.05 μs and the aperture is set to f/2.8. Since the FeO melting stage is much shorter than the Fe melting stage, frame rates of at least 80,000 fps are used, and up to 200,000 fps when required by the operating condition. The corresponding image resolutions, frame rates, and fields of view are listed in Table 3. The lower edge of the field of view is aligned with the top of the central tube.

Table 3
Frame rate, resolution, and field of view of the single-color camera under different experimental conditions.

| $T_{gas}$, K | Parameters | Oxygen concentration, % | | | | |
| --- | --- | --- | --- | --- | --- | --- |
| | | 11 | 21 | 31 | 41 | 51 |
| 1200 K | Frame rate | 80, 000 fps | | | 120, 000 fps | |
| | Resolution | 1024 pixels × 32 pixels | | | | |
| | FOV | 61.4 mm × 1.9 mm | | | | |
| 1500 K & 1800 K | Frame rate | 200, 000 fps | | | | |
| | Resolution | 512 pixels × 32 pixels | | | | |
| | FOV | 20.5 mm × 1.3 mm | | | | |

In the present experiments, the recorded blackbody radiation can be converted to particle temperature after calibration by using a single-color pyrometry. The detected radiation intensity from a particle is given by:

$$I = A\Omega\varepsilon_p t_{exp} \int R_{cam}(\lambda) Tr_{opt}(\lambda) L_{bb}(\lambda, T_p) d\lambda \quad (5)$$

where $A$ is the emitting area of the particle, $\Omega$ is the geometric collection factor, $R_{cam}$ is the spectral response of the camera, $Tr_{opt}$ is the spectral transmittance of the optics, $\varepsilon_p$ is the particle spectral emissivity, $L_{bb}$ is the blackbody spectral radiance, and $t_{exp}$ is the exposure time.

Temperature evaluation is restricted to the range between the FeO and Fe melting points (1652–1809 K) [35]. $\Omega$ is treated as constant. In addition, the outer $Fe_3O_4$ layer does not melt in the investigated temperature interval, so $\varepsilon_p$ is assumed to be a constant value of 0.88 [36] for simplicity, and the same value was used in the subsequent numerical simulations. As shown later, the measured FeO melting temperature supports the validity of these assumptions.

Since the melting point of iron is fixed, the luminosity at the instant of melting for a given particle can be used as a calibration point, so that:

$$\frac{I}{I_{Fe}} = \frac{A}{A_{Fe}} \frac{\int R_{cam}(\lambda) Tr_{opt}(\lambda) L_{bb}(\lambda, T_p) d\lambda}{\int R_{cam}(\lambda) Tr_{opt}(\lambda) L_{bb}(\lambda, T_{p,Fe}) d\lambda} \quad (6)$$

Neglecting the effect of the nanometer-scale oxide layer growth on the particle diameter during the investigated temperature range, the ratio $A/A_{Fe}$ can be obtained from the known density–temperature relation of iron [37] for the particles closest to a sphere. The second factor in Eq. (6) can also be calculated once $R_{cam}$ and $Tr_{opt}$ are known (Fig. 6). The transmittance of the long-pass filter and the spectral response of the high-speed camera are provided by the manufacturers, and the lens transmittance is measured with a spectrophotometer (Lambda 950, PerkinElmer). A lookup table relating $I/I_{Fe}$ to $T_p$ is then constructed, and the particle temperature $T_p$ is obtained from the measured $I/I_{Fe}$ using this table (Fig. 7).

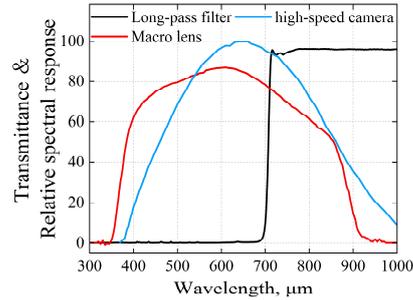

Fig. 6. Spectral transmittance of the long-pass filter and macro lens, and spectral response of the high-speed camera used for single-color pyrometry.

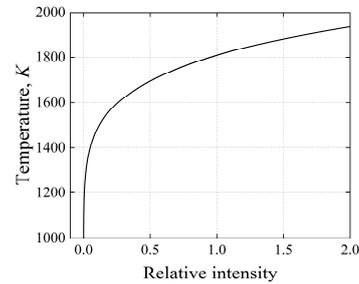

Fig. 7. Calibrated relation between relative single-color signal intensity ($I/I_{Fe}$) and particle temperature based on the iron melting point.

Using the calibrated single-color pyrometry, the FeO melting temperatures of individual particles were measured, as shown in Fig. 8. The mean FeO melting temperature is about 1645 ± 25 K, in good agreement with the nominal FeO melting point of 1652 K. This directly confirms the reliability of the single-color pyrometry in the present experiments.

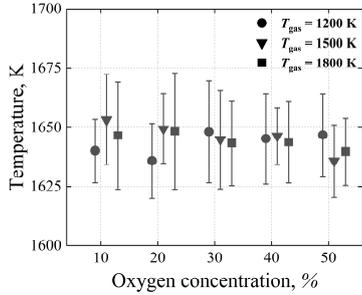

Fig. 8. Mean FeO melting temperatures of the particles with roundness greater than 0.9 from single-color pyrometry.

After the camera records the particle emission, a second-order polynomial fit to the particle trajectory is used to determine the reference time ($t_{py}$) at which the iron particle enters the burner. This entry time is then matched to the corresponding particle captured by the holography camera, which provides the actual entry time ($t_H$) into the burner, enabling a one-to-one association between particle diameter, morphology, the actual burner-entry time and its combustion time history. All data processing is performed using in-house MATLAB codes.

## 3. Modeling methodology

Numerical simulations are performed to describe the solid-phase oxidation of the iron particles. The model is based on solid-phase iron oxidation kinetics following a parabolic rate law intended to analyze iron particles' ignition temperature, whose details can be found in Mi et al. [7], and the same framework was also adopted in ref. [3, 38]. Only a brief model outline with the boundary and initial conditions is given here.

The model setup is illustrated in Fig. 9. In this ignition model, the particle consists of an outer $Fe_3O_4$ layer and an inner FeO layer. A two-stage oxidation mechanism is adopted. In the first stage, the initial oxidation rate is governed by a parabolic rate law for oxide-scale growth, which provides the only source of chemical heat release. In this regime, the oxygen consumption rate of the particle is lower than the maximum external-oxygen-transport rate, corresponding to the yellow region in Fig. 9. The oxide growth rate in this stage is given by [7]:

$$\frac{dX_i}{dt} = k_{0,i} \text{Exp}\left(\frac{-T_{a,i}}{T_p}\right)\frac{1}{X_i} \quad (7)$$

where $i$ denotes either the FeO or $Fe_3O_4$ oxide layer, $X_i$ is the oxide thickness, $k_{0,i}$ and $T_{a,i}$ are the pre-exponential factor and activation temperature, respectively, with values taken from Ref. [7]. Multiplying the oxide growth rate by the corresponding oxide density, particle surface area, and formation enthalpy yields the chemical heat release rate. Accounting for convective and radiative heat exchange with the surrounding gas gives the particle enthalpy change, and the particle temperature at each time step is obtained from the total enthalpy–temperature relation. The phase transitions are treated through the piecewise enthalpy–temperature relation of the particle materials, with the enthalpy discontinuity accounting for latent heat; therefore, no separate phase-transition-rate equation is required.

As seen from Eq. (7), the oxidation rate in this kinetics-controlled-regime is independent of the ambient oxygen concentration. Thus, for iron particles with same properties, both the ignition temperature and the solid-oxidation-time predicted by this model do not depend on oxygen concentration in this regime, and the ignition temperature independence has been confirmed experimentally [3, 5, 6].

When the oxygen consumption rate determined by parabolic rate law exceeds the maximum oxygen flow rate to the particle surface supplied by external transport, while the FeO layer is still solid, the particle enters the second stage, in which the reaction rate is controlled by the maximum external-oxygen-transport rate. Once FeO melts, the parabolic rate law is no longer valid, and the oxidation rate is also taken to be controlled directly by external-oxygen-transport.

The blue region in Fig. 9 corresponds to this external-oxygen-transport-controlled stage. As illustrated in the enlarged schematic of the particle in Fig. 9, oxygen at the particle surface is assumed to be consumed instantaneously, i.e., adsorption at the surface, diffusion through the oxide, and reaction with Fe or iron oxides are treated as infinitely fast. The oxygen concentration at the particle surface is therefore zero, and the maximum external-oxygen-diffusion rate can be written as [7]:

$$\dot{m}_{D,\max} = A_p \frac{\text{Sh} D_{O_2}}{d_p} C_{O_2,g} \quad (8)$$

where $A_p$ is the particle surface area, Sh is the Sherwood number, $D_{O_2}$ is the diffusivity of $O_2$ in the ambient gas, $d_p$ is the particle diameter, and $C_{O_2,g}$ is the $O_2$ concentration in the bulk gas. From Eq. (8), the oxidation rate in this stage is proportional to $C_{O_2,g}$.

Since oxygen at the particle surface is completely consumed and no other gas species are produced, there is a net gas consumption at the surface. In addition to diffusion, this drives gas toward the particle surface and induces Stefan flow, which increases the oxygen supply to the surface compared with pure diffusion (Eq. (8)). When Stefan flow is taken into account, the Sherwood number can be rewritten as [39]:

$$\text{Sh}_{\text{Stefan}} = \text{Sh} \cdot \frac{\ln(1 + B_m)}{B_m} \quad (9)$$

where $B_m$ is the Spalding mass transfer number [39], which reduces to $-C_{O_2,g}$ when the oxygen concentration at the particle surface is assumed at 1200 K and 11% $C_{O_2,g}$. The particle oxidation is divided into kinetic-controlled (yellow) and external-oxygen-transport-controlled (blue) regions.

Fig. 9. Schematic of the simulation model and calculated gas/particle temperature histories for a 38 μm iron particle

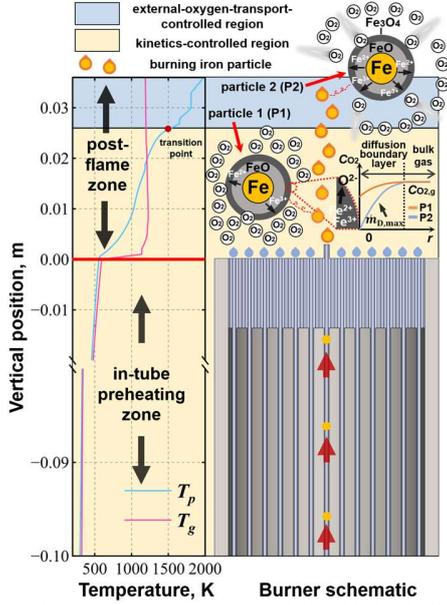

to be zero. The model neglecting Stefan flow, based on Eqs. (7) and (8), is denoted diffusion model, while the model including Stefan flow corrections via Eqs. (7) and (9) is denoted Stefan-flow model. In diffusion model, Sh follows the Frössling correlation. In Stefan flow, Sh and the corresponding Nusselt number (Nu) are taken from the boundary-layer-resolved model corrected correlations developed by Thijs et al. [23] for iron particle combustion. Since the particle temperatures considered here are relatively low, evaporation of iron and the associated formation of nano-sized oxides are neglected.

The gas-phase boundary conditions are specified using the equilibrium composition of complete methane combustion listed in Table 2, while the gas temperature is prescribed according to Fig. 1(b-d). Iron particles are preheated in the central tube before entering the high-temperature gas environment. If this preheating process is neglected and the initial temperature of particles entering the high-temperature gas is simply set to ambient, the $SOT_{FeO}$ of a 30 μm iron particle at a $T_g$ of 1200 K would increase by more than 15%, which would compromise the accuracy of the simulation. To account for the thermal history of particles prior to entering the post-flame gas region, the environmental temperature in the in-tube preheating zone shown in Fig. 9 is considered. It is assumed that heat conduction in the honeycomb mesh region on the burner surface occurs only along the gas-flow direction, such that the problem can be reduced to one-dimensional conduction, and that in the central tube only axial heat conduction and convection are present. The governing equations for the wall temperature of the central tube and the gas-phase temperature are given by:

$$k_s A_s \frac{d^2 T_s}{dx^2} = (h_{in} P_{in} + h_{out} P_{out})(T_s - T_g) \quad (10)$$

$$\dot{m} c_p \frac{dT_f}{dx} = (h_{in} P_{in} + h_{out} P_{out})(T_s - T_g) \quad (11)$$

where $k_s$ is the thermal conductivity of the central tube wall, $A_s$ is the cross-sectional area, $T_s$ is the wall temperature, $T_g$ is the gas temperature, and $P_{in}$ and $P_{out}$ are the inner and outer wetted perimeters. The gas mass flow rate is $\dot{m}$, and $c_p$ is the gas specific heat.

To simplify the problem, the gas temperature around the tube and mass flux are assumed to be the same inside and outside the central tube (taken as the internal values), so that only one gas energy equation is needed to solve. The inner and outer convective heat transfer coefficients, $h_{in}$ and $h_{out}$, are obtained from $h = $ Nu $* k_g / L_c$, where $k_g$ is the gas thermal conductivity, and $L$c is a characteristic length. The Nusselt numbers inside and outside the central tube are given by [40,41]:

$$Nu_{in} = 3.66 + \frac{0.0668 Re_{in} \Pr\left(\frac{D_i}{L}\right)}{1 + 0.04\left(\frac{Re_{in} \Pr D_i}{L}\right)^{\frac{2}{3}}} \quad (12)$$

$$Nu_{out} = 0.664 Re_{out}^{\frac{1}{2}} \Pr^{\frac{1}{3}} \quad (13)$$

where $Re_{in}$ and $Re_{out}$ are the Reynolds numbers of the gas inside and outside the capillary, respectively. As shown in Fig. 9, the length of the central tube below the burner surface is 0.1 m, and its lower end corresponds to the initial position of the iron particles. The inner and outer diameters of the tube are 0.5 mm and 0.7 mm, respectively. The inlet gas and particle temperature are set to 300 K, while the outlet gas temperature is specified as the burner surface temperature given in Table 2. By solving Eqs. (10) and (11), the gas temperature distribution inside the central tube can be obtained. The gas velocity is calculated from the ideal gas equation of state. The acceleration of an iron particle is determined by:

$$m_p \frac{d(u_p - u_g)}{dt} = (\rho_p - \rho_g) V_p g - F_D - m_p \frac{du}{dt} \quad (14)$$

where $m_p$ is the particle mass, $u_p$ is the particle velocity, $u_g$ is the gas velocity, $\rho_p$ and $\rho_g$ are the particle and gas densities, $V_p$ is the particle volume, $g$ is the gravitational acceleration, and $F_D$ is the Stokes drag on the particle:

$$F_D = 3\pi \mu d_p (u_p - u_g) \quad (15)$$

where μ is the gas dynamic viscosity and $d_p$ is the particle diameter. The initial particle velocity is set equal to the gas velocity, and Eq. (14) is integrated to obtain the particle trajectory. In all simulations, iron particles are assumed to be spherical. The numerical calculations are performed using in-house MATLAB codes.

## 4. Iron particle

The iron particles used in the experiments were the same as those in our previous ignition-temperature study [3]: solid, nearly spherical particles with most diameters in the range 17–45 μm and a purity of 99.3%. Their average initial oxide layer thickness was 70 nm [3], as measured by an oxygen analyzer. Since the particle surface appeared silver-gray, the presence of red $Fe_2O_3$ in the oxide layer was considered negligible. In addition, because FeO is unstable under room-temperature storage conditions, the initial oxide layer was assumed to consist mainly of $Fe_3O_4$. Specifically, in the calculations, the initial FeO thickness was set to only 5% of the total oxide-layer thickness to initialize the model. Moreover, under the parabolic rate law, the heat released by the oxidation of Fe to FeO accounts for more than 90% of the total chemical heat release. Therefore, even if the total initial oxide layer thickness varies, the initial FeO thickness remains in nano-scale, and its influence on the overall combustion process is expected to be limited.

## 5. Results and discussions

In this subsection, representative experimental results for the luminosity and temperature evolution of iron particles near the iron melting point are first presented together with the corresponding simulations. Subsequently, the durations and onset times of the characteristic temperature plateaus are shown and analyzed based on both experimental data and simulation results.

### 5.1 Combustion process overview

Fig. 10(a) shows a representative iron particle captured and reconstructed by the holography camera, together with its subsequent luminosity evolution recorded by the pyrometry camera. The particle has an equivalent diameter of 38 μm, and the time when it is first captured by the holography camera, i.e., when it enters the burner, is defined as $t_0 = 0$ ms. After 19 ms, the particle is first detected by the pyrometry camera, then reaches a first luminosity (temperature) plateau $t_{12}$, followed by a shorter second plateau $t_{34}$, and finally a longest third plateau $t_{56}$. As $C_{O2}$ increases, all characteristic times become shorter, and the shortest ones, $t_{34}$ and $t_{12}$, may fall below the temporal resolution of the camera. For example, at gas temperature of 1200 K, the occurrence frequencies of $t_{34}$ and $t_{12}$ decrease from 83% and 96% at 11% $C_{O2}$, respectively, to 23% and 81% at 51% $C_{O2}$. When a plateau is present, the second derivative of luminosity with respect to time attains a minimum at its onset and a maximum at its end, which is used to identify plateau onset times and durations in data processing.

Fig. 10(b) shows the temperature history of the same particle as in Fig. 10(a), measured by single-color pyrometry, together with the simulated temperature history for a particle of the same diameter under identical conditions. The oxidation region switches from kinetics-controlled to external-oxygen-transport-controlled at about 20.7 ms. Both experiment and simulation exhibit three temperature plateaus. In the single-color calibration, $t_{56}$ is assumed to correspond to Fe melting and is used as the reference point. With this calibration, the experimental and simulated $t_{12}$ and $t_{56}$ agree closely in duration, onset time, and temperature. Thus, $t_{12}$ and $t_{56}$ can be interpreted as the FeO-scale melting and Fe melting stages, respectively. In contrast, the experimental and simulated $t_{34}$ differ significantly in onset time and temperature. In the simulations, $t_{34}$ is associated with the γ-Fe to δ-Fe phase transition.

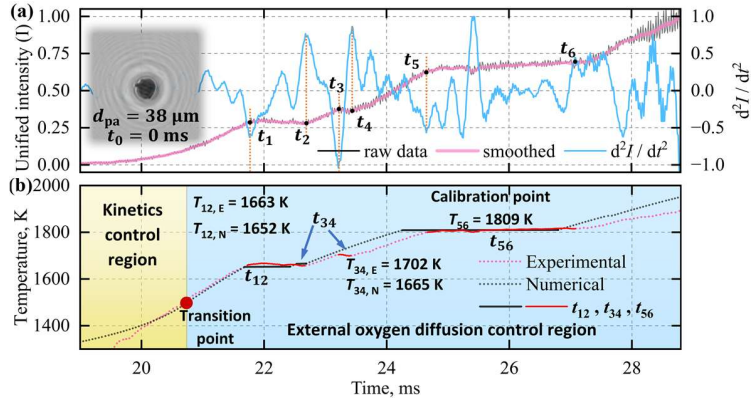

Fig. 10. (a) Reconstructed holographic image and temporal evolution of raw luminosity, smoothed luminosity, and its second derivative for a single iron particle ($d_{eq}$ = 38 μm) at 1200 K and 11% $C_{O2}$. The entry time $t_0$ is set to 0 ms; the first, second, and third luminosity plateaus are labeled $t_{12}$, $t_{34}$, and $t_{56}$. (b) Temperature history of the same particle from single-color pyrometry and from simulation. The simulation time is shifted so that the onset of FeO melting coincides with the experiments, facilitating comparison. $T_N$ and $T_E$ denote the simulated and experimental temperatures, respectively.

Section 5.3 shows that the experimental $t_{34}$ also corresponds to a γ-Fe to δ-Fe transition, but with an elevated transition temperature, possibly due to impurities. The measured particle temperatures below 1400 K deviate from the simulated values because the particle signal is weak in this range. At around 1400 K, it is only 5% of the full-well intensity, leading to a low signal-to-noise ratio. In addition, as shown in Fig. 7, the inferred temperature is more sensitive to signal fluctuations at low temperatures, resulting in larger measurement uncertainty.

*5.2 FeO melting onset and duration*

Based on Fig. 10, the oxidation time before FeO scale melting is denoted $t_{01}$, i.e., $SOT_{FeO}$. Fig. 11 shows $t_{01}$ for particles with different diameters and roundness. The roundness is defined as:

$$R = \frac{4\pi \cdot Area}{Perimeter^2} \quad (13)$$

When the particle cross-section is close to circular, the roundness approaches 1; as it deviates from a circle, the roundness approaches 0. As seen in Fig. 11, for particles that deviate from a spherical shape, $SOT_{FeO}$ tends to be shorter than that of particles with the same projected-area equivalent diameter. This is readily understood: non-spherical particles have a larger surface area, which increases the oxidation heat release rate proportionally, whereas radiative and convective heat losses do not increase in the same proportion. As a result, these irregular particles ignite more readily and exhibit shorter $SOT_{FeO}$.

Fig. 12 presents the diameter-binned mean $SOT_{FeO}$ for particles with roundness greater than 0.9, together with the corresponding simulations. Only particles with $R>0.9$ are used, since for irregular particles the projected-area equivalent diameter introduces errors in surface area and volume and thus in comparison with the models. The binned $SOT_{FeO}$ values clearly show the weak dependence on oxygen concentration. The simulations agree with the measurements, except for a slight overprediction for particles larger than 30 μm at 1200 K. This discrepancy is most likely due to an underestimated central tube temperature near the burner surface: the burner temperature is measured on the stainless-steel honeycomb rather than on the tube itself, and radial heat conduction loss from the honeycomb can make the actual tube temperature exhibits only a very weak dependence on oxygen concentration. This is because, as illustrated in Fig. 13, even at a low oxygen concentration of 11% and a gas temperature of 1200 K, the kinetics-controlled oxidation time ($t_k$) of 10–55 μm iron particles accounts for about 97% of $SOT_{FeO}$, i.e., the oxidation process before FeO melting follows the parabolic rate law for almost the entire duration.

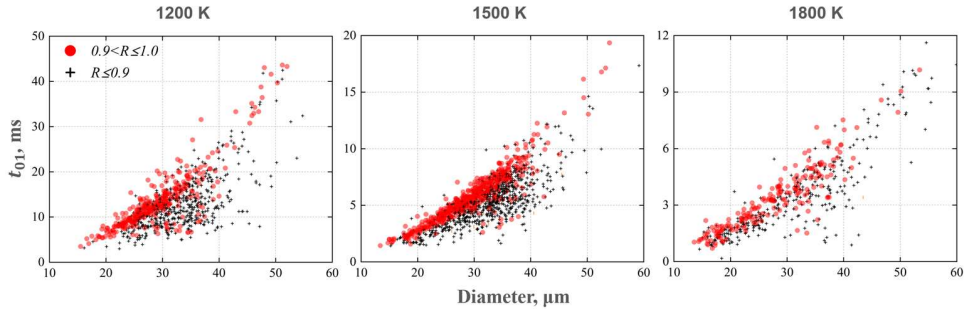

Fig. 11. $t_{01}$ for particles with different diameters and roundness ($R$). Here $t_{01}$ is the time from burner entry to the onset of FeO-scale melting, i.e., $SOT_{FeO}$.

Moreover, Fig. 14 shows that at higher oxygen concentrations (e.g., 31% and above), the oxidation rate does not become external-oxygen-transport-controlled before FeO melting, and the parabolic rate law for scale growth is essentially independent of oxygen concentration. Consequently, both the simulations and the experiments show that $SOT_{FeO}$ is insensitive to oxygen concentration.

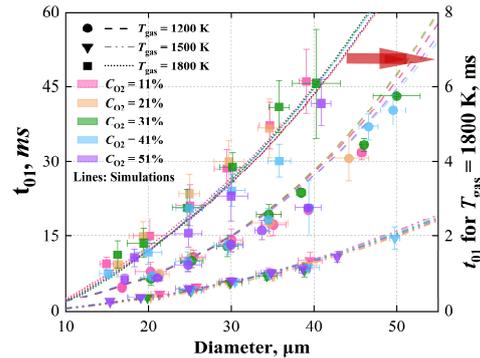

Fig.12. Diameter-binned mean $t_{01}$ for particles with roundness (R > 0.9) from experiments, together with the corresponding simulation results.

The values of $SOT_{FeO}$ reported in this subsection are simulated using the Stefan-flow model. Although Stefan flow can increase the maximum oxygen flow rate by about 40% at 51% oxygen, thereby delaying the transition point in Fig. 10(b) and formally extending $t_k$, the rapid particle heating leads to an exponential increase of the scale-growth rate—and thus the oxygen consumption rate—with particle temperature (Eq. (7)), so that the resulting extension of $t_k$ can be neglected.

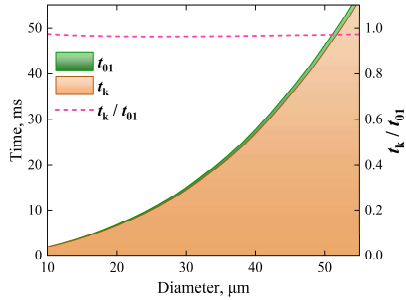

Fig. 13. Kinetics-controlled oxidation time ($t_k$), $t_{01}$, and their ratio for particles of different diameters at 1200 K and 11% $O_2$, calculated using the Stefan-flow model.

The above comparison between experiments and simulations strongly supports the parabolic rate law model for describing the oxidation of micron-sized iron particles, as long as the oxygen consumption rate remains below the maximum external-oxygen-transport rate. Nguyen et al. [11] refitted the kinetic parameters to match the experiments when using the first-order surface kinetics model. However, such case-by-case adjustment of kinetic parameters increases both experimental and modeling effort, and leads to ignition temperatures that depend on the chosen parameters. Moreover, the ignition temperature predicted by this model depends on the oxygen concentration. In contrast, the parabolic-rate-law-based model predicts ignition temperatures whose values and insensitivity to oxygen concentration are consistent with experiments [3,5,6], making it a simpler and more reliable framework for modeling iron particle ignition and solid-phase oxidation.

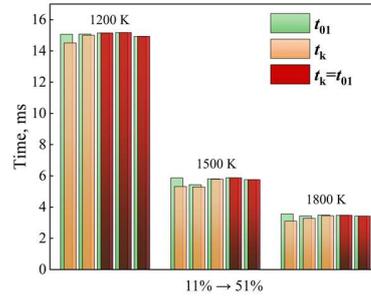

Fig. 14. $t_k$ and $t_{01}$ of 30 μm iron particles at various $O_2$ concentrations and temperatures from Stefan-flow model.

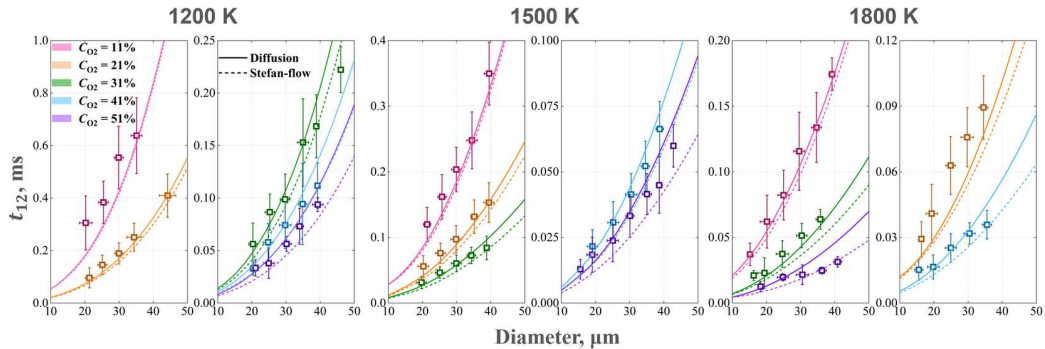

Fig. 15. FeO melting time $t_{12}$ for particles with different diameters. Hollow symbols denote diameter-binned means for particles with $R > 0.9$ and lines denote simulations results.

Fig. 15 shows the binned averages $t_{12}$ for particles with roundness larger than 0.9 and the corresponding simulations based on the diffusion and Stefan-flow models. The binned averages are reasonably captured by both the diffusion and Stefan-flow models at $C_{O2}$ up to 31%. At 41% and 51% $C_{O2}$, the experimental $t_{12}$ values lie between the two model predictions, with the diffusion model giving longer $t_{12}$ values and the Stefan-flow model giving shorter ones. For example, at 51% $C_{O2}$, the Stefan-flow model predicts a $t_{12}$ 30% shorter than that predicted by the diffusion model, and the binned mean experimental values remain within this range. The agreement between experimental and simulated $t_{12}$ further supports the assumption that, once the particle temperature reaches the FeO melting point, the oxidation rate is limited by external-oxygen-transport.

### 5.3 Onset and duration of the γ-Fe to δ-Fe phase transition

Fig. 16 shows the diameter-binned experimental durations of the second luminosity (temperature) plateau, $t_{34}$, for particles with $R > 0.9$, together with the simulated duration of the γ-Fe to δ-Fe phase transition. Since $t_{34}$ is short, fewer valid data points are available and the temporal-sampling uncertainty is larger than for the other characteristic times. Most experimental points lie between the two model predictions, supporting the assignment of $t_{34}$ as the γ-Fe to δ-Fe phase transition, although larger deviations appear for the 1200 K data at 11% and 21% $C_{O2}$, likely due to the limited number of valid measurements. For small particles at higher $C_{O2}$, the deviations are also more noticeable, which is likely related to the very short duration of $t_{34}$.

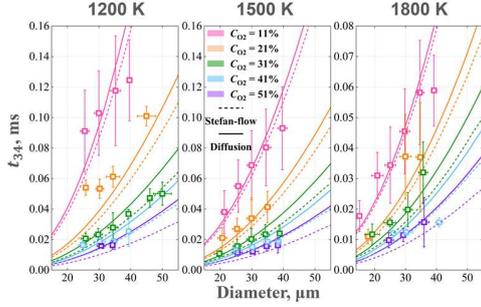

Fig, 16. Diameter-binned mean $t_{34}$ for particles with $R > 0.9$, together with simulated durations of the second luminosity (temperature) plateau.

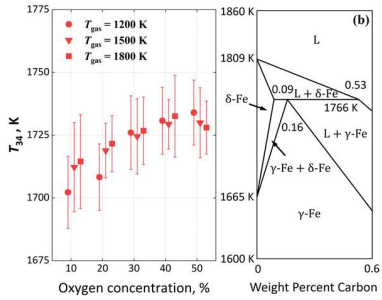

Fig. 17. (a) Diameter-binned mean $T_{34}$ for particles with $R > 0.9$; (b) Fe–C phase diagram, adapted from Ref. [42].

Fig. 17(a) presents the mean temperature during the $t_{34}$ interval, $T_{34}$, measured by single-color pyrometry. The values lie between 1700 and 1730 K, significantly higher than the γ-Fe to δ-Fe transition temperature of 1665 K for pure iron [35]. This shift is likely caused by impurities such as carbon. As shown in the Fe–C phase diagram in Fig. 17(b), a carbon content of 0.09 wt% raises the γ-Fe to δ-Fe transition temperature from 1665 K to 1766 K, and the measured $T_{34}$ falls within this range. Moreover, Fig. 17(a) shows that $T_{34}$ increases with ambient oxygen concentration. As the present experiments provide only a single parameter—the γ-Fe to δ-Fe transition temperature—a detailed explanation of this trend must be left to a more in-depth studies in the future. In summary, the $t_{34}$ duration corresponds to the γ-Fe to δ-Fe phase transition, but its temperature is elevated above the ideal value by impurities.

### 5.4 Fe melting onset and duration

Fig. 18 shows the diameter-binned mean $t_{25}$ and the corresponding simulations, where $t_{25}$ is the time between the end of the first and the start of the third luminosity (temperature) plateau. The value of $t_{25}$ decreases with increasing $C_{O2}$ and $T_g$, and agrees better with the diffusion model predictions, whereas Stefan-flow model slightly underestimates $t_{25}$.

Fig. 19 shows the iron melting time $t_{56}$ for particles with different roundness, together with diameter-binned mean $t_{56}$ for particles with ($R > 0.9$) and the simulated Fe melting times. Overall, the experimental and simulated $t_{56}$ values agree reasonably well. The trends in the comparisons between them are similar to those for $t_{12}$ and $t_{34}$: reasonable agreement is obtained at low O₂, while at higher oxygen levels the experimental data tend to lie between the two model predictions.

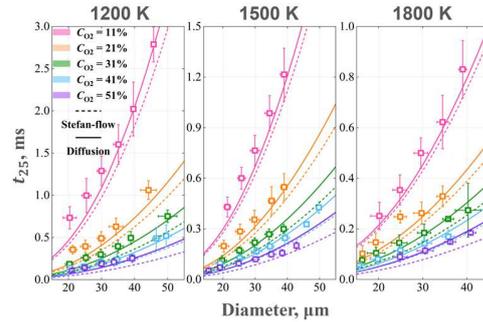

Fig. 18. Diameter-binned mean $t_{25}$ from experiments (particles with $R > 0.9$) and simulations; $t_{25}$ is the time between the end of the first and the start of the third luminosity (temperature) plateau.

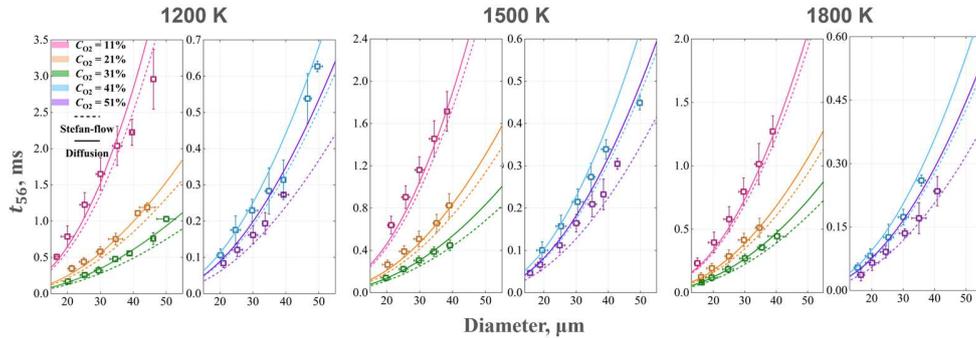

Fig. 19. Diameter-binned mean $t_{56}$ for particles with roundness ($R > 0.9$), and corresponding simulated durations of the third luminosity (temperature) plateau.

In this study, SOT ($t_{05}$) are not shown explicitly, because $t_{05}$ is simply the sum of the already analyzed intervals $t_{01}$, $t_{12}$, and $t_{25}$; their reasonable agreement with simulations imply similarly agreement for $t_{05}$. The model used by Nguyen et al. [11] is essentially equivalent to the present diffusion model, but in their work the simulated SOTs are systematically lower than the experimental results [10]. Awad [14] extended a model similar to Stefan-flow model by including Knudsen-transition transport [15,16] and surface chemisorption of O₂; their predictions are closer in magnitude to the SOT data of Ning et al. [10], but still fail to capture the trend with oxygen concentration. Further investigation with necessary replicability verification is required to explain this discrepancy.

Overall, despite remaining discrepancies for the γ-Fe to δ-Fe transition temperature and onset, the other characteristic times—$SOT_{FeO}$, FeO melting, γ-Fe to δ-Fe transition duration, the interval between FeO and Fe melting, and Fe melting—are well reproduced by the simulations. This supports the parabolic rate law model for solid-phase oxidation of iron and confirms that, between FeO and Fe melting, the oxidation rate is limited by external-oxygen-transport.

## 6. Conclusions

High-resolution in-line holography and ultra–high-speed single-color pyrometry were used to measure characteristic oxidation times of well-defined single iron particles near their melting point, with accurately resolved initial diameters. The data show three distinct temperature plateaus: FeO scale melting, the γ-Fe to δ-Fe phase transition, and Fe melting. The oxidation time before FeO melting is essentially insensitive to oxygen concentration, whereas the durations of the plateaus and the intervals between them decrease with increasing oxygen concentration once FeO starts to melt.

A solid-phase oxidation model combining a parabolic rate law with an external-oxygen-transport-controlled description accurately reproduces all characteristic times in the solid-phase stage. This provides time-resolved validation that the parabolic rate law reliably describes iron oxidation before FeO melting, while after FeO melting the rate is controlled by external-oxygen-transport. The results offer stringent experimental support for existing pre-melting iron combustion models and a robust basis for model selection and validation in future simulations.

## CrediT authorship contribution statement

• Liulin Cen: Conceptualization, Methodology, Investigation, Data curation, Visualization, Formal analysis, Writing – original draft.
• Yong Qian: Supervision, Methodology, Writing – review & editing, Project administration.
• XiaoCheng Mi: Visualization, Formal analysis, Methodology, Writing – review & editing.
• Xingcai Lu: Supervision, Resources, Writing – review & editing, Project administration.

## Declaration of competing interest

The authors declare that they have no known competing financial interests or personal relationships that could have appeared to influence the work reported in this paper.

## Acknowledgments

This research was funded by the National Natural Science Foundation of China under Grant Nos. 52276128 and 52236007. LC also acknowledges financial support from the China Scholarship Council.